\documentclass[aps,prc,twocolumn,superscriptaddress]{revtex4-1}

\usepackage{amsmath}
\usepackage{graphicx}
\usepackage{units}
\usepackage{url}

\begin{document}


\title{Study of the sensitivity of observables to hot spot size in heavy ion collisions}



\author{Fernando G. Gardim}
\affiliation{Instituto de Ci\^encia e Tecnologia, Universidade Federal de Alfenas, 37715-400 Po\c{c}os de Caldas-MG, Brazil}

\author{Fr\'ed\'erique Grassi}
\affiliation{Instituto de F\'{i}sica, Universidade de S\~ao Paulo, Rua do Matão 
1371,  05508-090 S\~ao Paulo-SP, Brazil}

\author{Pedro Ishida}
\affiliation{Instituto de F\'{i}sica, Universidade de S\~ao Paulo, Rua do Matão 
1371,  05508-090 S\~ao Paulo-SP, Brazil}

\author{Matthew Luzum}
\affiliation{Instituto de F\'{i}sica, Universidade de S\~ao Paulo, Rua do Matão 
1371,  05508-090 S\~ao Paulo-SP, Brazil}

\author{Pablo S. Magalh\~aes}
\affiliation{Instituto de Ci\^encia e Tecnologia, Universidade Federal de Alfenas, 37715-400 Po\c{c}os de Caldas-MG, Brazil}


\author{Jacquelyn Noronha-Hostler}
\affiliation{Department of Physics and Astronomy,
Rutgers, The State University of New Jersey, Piscataway, NJ 08854-8019, USA}


\date{\today}

\begin{abstract}
An open question in the field of heavy-ion collisions is to what extent the size of initial inhomogeneities in the system affects measured observables.  
Here we present a method to smooth out these inhomogeneities 
with  minimal effect on global properties, in order to quantify the effect
of short-range features of the initial state.
We show a comparison of hydrodynamic predictions with original and smoothened initial conditions for four models of initial conditions and various observables.
Integrated observables (integrated $v_n$, scaled $v_n$ distributions, normalized symmetric cumulants, event-plane correlations) as well as most differential observables ($v_n(p_T)$) show little dependence on the inhomogeneity sizes, and instead are sensitive only to the largest-scale geometric structure. However other differential observables such as the flow factorization ratio and sub-leading principal components are more sensitive to the granularity and could be a good tool to probe the short-scale dynamics of the initial stages of a heavy-ion collision, which are not currently well understood.
\end{abstract}


\maketitle

\section{Introduction}

Relativistic heavy ion collisions are being performed at RHIC and the LHC to 
study the Quark Gluon Plasma. The aim is to extract its transport properties, phase diagram, and  
initial state.  Understanding its initial state, for instance, can help clarify details of strong interactions away from equilibrium. In the standard picture of a relativistic heavy ion collision,  the system
rapidly thermalizes and expands hydrodynamically (for recent reviews see \cite{reviewHeinz,reviewGale,reviewdeSouza,reviewQGP5}).  Ultimately the system decouples and particles are emitted. However, the initial stages of the collisions, before the system has sufficiently thermalized to exhibit hydrodynamic behavior, are still poorly understood.
Hydrodynamic simulations therefore rely on models to provide initial conditions,
of which many exists, with various features and levels of sophistication.
There are differences in the source of fluctuations in each of these different initial condition models, for instance, contributions of the quarks and gluons to fluctuations vs. assuming only nucleonic fluctuations, which translates into different scales of structure.

In models based on the Monte Carlo Glauber model \cite{glauber1,glauber2,glauber3}, 
nucleons follow straight-line trajectories and make collisions. In coordinate space the positions of the wounded nucleons are like delta function, thus, two-dimensional Gaussians are used to smear the colliding nucleons.  The usual source of
fluctuations is the position of the nucleons so  the size of the hot spots reflects roughly the radius of a proton ($\sim 1$ fm). More recently an alternative to the standard wounded nucleon picture was created using parameterized version of initial conditions, TRENTO \cite{Moreland:2014oya}. At this point in time, sub-nucleonic degrees of freedom have not yet be included in the public version. 

More sophisticated models with non-trivial dynamics are also employed such as NeXus \cite{NeXus}, EPOS \cite{Pierog:2013ria}, UrQMD \cite{Bleicher:1999xi,Bass:1998ca}, and AMPT \cite{Zhang:1999bd}. These can involve various scales: in the NeXus model \cite{NeXus}, parton
 ladders are exchanged between nucleons, 
fluctuations occur  both at the nucleonic level  - nucleon positions fluctuate - and partonic level - energy sharing  to produce the ladders is probabilistic but the hot spot size also reflects the nucleon size \cite{NeXIC}. This is illustrated in the first row of Fig.~\ref{fig:IC}.

Models based on perturbative QCD combined with saturation physics also exist, such as the EKRT model \cite{Niemi:2015qia}.
Finally, there are models based on the Color-Glass-Condensate effective theory, most notably MC-KLN \cite{MCKLN} and  IP-Glasma \cite{Schenke12}. 
In the MC-KLN model \cite{MCKLN}, at a certain point in the transverse plane (x,y) the energy density depends on the saturation scale, which is related to the nuclear thickness functions through the $k_t$-factorization formula.
Nucleonic fluctuations are considered in mckln, although small uncorrelated hot spots  appear in certain versions, as shown in the bottom row  of Fig.~\ref{fig:IC}.
 In the IP-Glasma model \cite{Schenke12},  fluctuations of nucleon positions
as  well  as  sub-nucleonic  fluctuations  of  color  charges are included. The resulting hot spot size is significantly smaller \cite{Schenke12} than other models. 

Many of these models have been quite successful in reproducing experimental data (for a few recent comparisons see \cite{Shen:2015qta,Bernhard:2016tnd,Alba:2017hhe,Giacalone:2017dud,Eskola:2017bup,McDonald:2016vlt}).  However, each of these models have differences in the macroscale i.e. shape and size of the initial conditions, the size/location of the hot spots, and the strength of the fluctuations such that it is not always clear exactly which features are essential for reproducing a given observable.
In particular, many observables can be simultaneously reproduced by different initial condition models, providing the transport properties and other relevant parameters are properly adjusted. Significant work has been done in terms of constraining the degree of fluctuations in initial conditions using multi-particle cumulants \cite{Giacalone:2017dud} and event-by-event flow distributions \cite{Renk:2014jja}.  

One open question is whether the spatial extent of ``hot spots'' in the initial system --- which can be quite different in different models --- has a sizable effect on measured observables.
%
This is an important question if we want to rule out initial conditions models and elucidate the dynamics of the strong interactions. This has been studied previously \cite{Petersen:2010zt,ColemanSmith:2012ka,RihanHaque:2012wp,Bzdak:2013zma,Floerchinger:2013rya,Floerchinger:2013vua,Floerchinger:2013hza,Retinskaya:2013gca,Renk:2014jja,Konchakovski:2014fya} often using Gaussians to smooth out small scale fluctuations, which has been shown to quickly increase the radius of the eccentricities, thus, making the initial conditions rounder as one smooths out small scale structure \cite{Bhalerao:2011bp}.  More recently, cubic splines have been used \cite{Noronha-Hostler:2015coa}, which smooth out fluctuations to a finite radius, thus, preserving the initial eccentricities out to larger smoothing scales \cite{Noronha-Hostler:2015coa}.  An alternative approach has been recent work evolving extremely spiky initial conditions that produce large Knudsen numbers until they initial conditions reach a point where they are applicable for hydrodynamics \cite{Keegan:2016cpi}.

The objective of this paper is therefore the following: we consider four initial state models and smooth the size of their inhomogeneities on scale from 0.3 to 1 fm (we do not go farther since this is the typical range of nucleonic inhomogeneities). We then compare predictions for  observables  for the original model and its  smoothed versions. We find that the differences are in fact very small for a large range of observables: integrated $v_n$, scaled $v_n$ distributions, normalized symmetric cumulants, event plane correlations, and $v_n(p_T)$.  We observe larger differences  for the 
flow factorization  ratios $r_n$ and the sub-leading modes in a principal component analysis of the two-particle correlation matrix. 

The outline of this paper is as follows. In section II, we recall how to characterize the initial and final states through eccentricities and
harmonic flow coefficients. In section III, we  describe our smoothing method. In section IV, we present  our results for integrated and $p_T$ dependent quantities. Finally in section V, we discuss our findings.

\section{Characterizing the initial and final states}

The initial conditions for hydrodynamic evolution consist of the energy-momentum tensor at some initial time $T^{\mu\nu}(\tau = \tau_0)$.  It is believed that the most important aspect of the initial conditions for observables near mid-rapidity is the energy density in the transverse plane $\epsilon(x,y) = T^{\tau\tau}(\tau_0, \eta \sim 0, x,y)$.  
Hydrodynamic evolution converts this geometry into the final momentum distribution of detected particles.

We would like to characterize this initial density distribution in a way that is ordered according to length scale.  The natural way to do this is to switch to Fourier transformed coordinates, such that small $k$ represents large-scale structure and large $k$ represents small-scale structure.   
%
%
%
%

%

Specifically, we define the transformed density via a 2D Fourier transform \cite{teaney1}
\begin{equation}
\rho(\vec{k}) = \int d^2x \epsilon(\vec{x}) e^{-i \vec{k}\cdot \vec{x}} ,
\end{equation}
from which we create a cumulant generating function
\begin{equation}
e^{W(\vec k)}\equiv\rho({\vec{k}})  ,
\end{equation}
that we expand in a power series around $k \equiv |{\vec k}| = 0$:
\begin{align}
W(\vec k)  &= \sum_{m=0}^\infty W_m(\phi_k) k^m  .
\end{align}
It is useful to encode the dependence on azimuthal angle $\phi_k$ in a Fourier series, to obtain a discrete set of coefficients that contain all information about the distribution of energy density $\epsilon(x,y)$,
\begin{align}
W(\vec k)  = \sum_{m=0}^\infty \sum_{n=-\infty}^\infty W_{n,m} k^m e^{-i n\phi_k }.
\end{align}

The coefficients with smallest $m$, therefore, represent information about the largest-scale, global structure, while larger $m$ represents smaller-scale structures in the initial geometry.  The value of $n$ represents the rotational property of each coefficient.

Note that non-zero coefficients must have $m \geq n$, and $m-n$ must be even.  So for a given Fourier harmonic $n$, the lowest cumulant is $W_{n,n}$.

This expression can be inverted to obtain explicit equations for the coefficients $W_{n,m}$ (called cumulants, since they have the same relation to the distribution of energy as traditional cumulants have to a probability distribution).  We list a few of the lowest cumulants here, defining the complex coordinate $z \equiv x + i y$:
\begin{align}
W_{0,0} &= \ln E\\
W_{1,1} &\propto \langle  z\rangle\\
W_{0,2} &\propto   \langle  |z|^2 \rangle 
- \langle  z \rangle \langle  \bar z \rangle \\
W_{2,2} &\propto  \langle  z^2 \rangle - \langle  z\rangle^2 \\
W_{1,3} &\propto \langle  z^2 \bar z \rangle - \langle  z^2 \rangle \langle  \bar z \rangle  - 2\langle  |z|^2 \rangle \langle  z \rangle + 2 \langle  z \rangle^2 \langle  \bar z \rangle \\
W_{3,3} &\propto \langle  z^3 \rangle \nonumber + \langle  z \rangle \left (3 \langle  z^2 \rangle - 2\langle  z \rangle^2 \right),
\end{align}
with 
\begin{equation}
\langle  \ldots \rangle =  \frac{\int  d^2x \epsilon({\bf{x}}) \ldots} {\int  d^2x \epsilon({\bf{x}})} ,
\end{equation}
and $E$ is the total energy $E = \int d^2x \epsilon({\bf x})$.

With this construction, all cumulants are invariant under translations, except $W_{1,1}$, which represents the center of the system.  They are therefore appropriate for making a connection to the final momentum-space particle distributions, which do not depend on the choice of coordinate center.

To study dimensionless observables such as anisotropic flow, it is useful to define dimensionless ratios out of the lowest cumulants for each azimuthal harmonic $n$, whose magnitude and phase are the standard eccentricities $\varepsilon_n$ and participant planes $\Phi_n$:
\begin{align}
\mathcal{E}_2  = \varepsilon_2 e^{2i\Phi_n} &\equiv -2 \frac {W_{2,2}}{W_{0,2}} = -\frac{\langle  z^2\rangle}{\langle  |z|^2\rangle} = -\frac{\langle  r^2 e^{2i\phi}\rangle}{\langle  r^2\rangle}\\
\mathcal{E}_3 &\equiv -\frac{\langle  r^3 e^{3i\phi}\rangle}{\langle  r^3\rangle}\\
\mathcal{E}_1 &\equiv -\frac{\langle  r^3 e^{i\phi}\rangle}{\langle  r^3\rangle},
\end{align}
etc, where it is understood that the center of coordinates is chosen in each event such that $W_{1,1}=0$, which significantly simplifies the expressions.  

If we also expand the final single-particle momentum distribution in an azimuthal Fourier series,
\begin{align}
\frac{dN}{d\phi_p} = \frac{N}{2\pi}\sum_n V_n e^{-in\phi_p}
\end{align}
with
\begin{equation}
V_n \equiv v_n e^{in\Psi_n}=\frac 1 N\int d\phi_p e^{in\phi_p}\frac{dN}{d\phi_p},
\end{equation}
(Differential $V_n(p_T,\eta)$ can be defined in a similar way.)\\
We can conjecture event-by-event vector relations such as
\begin{align}
V_2 &\propto \mathcal{E}_2,\nonumber\\
V_3 &\propto \mathcal{E}_3.
\label{linear}
\end{align}
It has been shown that these relations are quite accurate, on an event-by-event basis \cite{Niemi12,Niemi15,Fu15} and for differential measurements as well \cite{Noronha-Hostler:2016eow,Betz:2016ayq,Prado:2016szr}.  

This is a very deep statement about the nature of hydrodynamic behavior --- the eccentricities $\varepsilon_n$ represent  only the lowest in an infinite series of cumulants with harmonic $n$, representing global properties at the largest length scales.  Even in cases where a non-linear dependence on eccentricities is known (such as $v_4$ and $v_5$ in non-central collisions), the fact that it depends only on eccentricities $\mathcal{E}_n$ still indicates that the final observables are dominated by structures in the initial energy density at the longest length scales.

It is therefore known that momentum integrated, as well as differential, flow depends mostly on the largest length scales, as represented by eccentricities $\varepsilon_n$.  However, the above relations are not 100\% precise, and there is room for some sensitivity to structures in the initial state at smaller length scales.

In this work, we investigate this possible sensitivity to the granularity of the initial energy density profile in the transverse plane, and want to find observables that can best probe these features.  To do this, we must establish a dependence of these observables on higher cumulants $W_{n,m}$, with $m>n$, beyond any dependence on eccentricities, which only represent global properties.

\section{Smoothing method}

\begin{figure*}
\includegraphics[width=\linewidth]{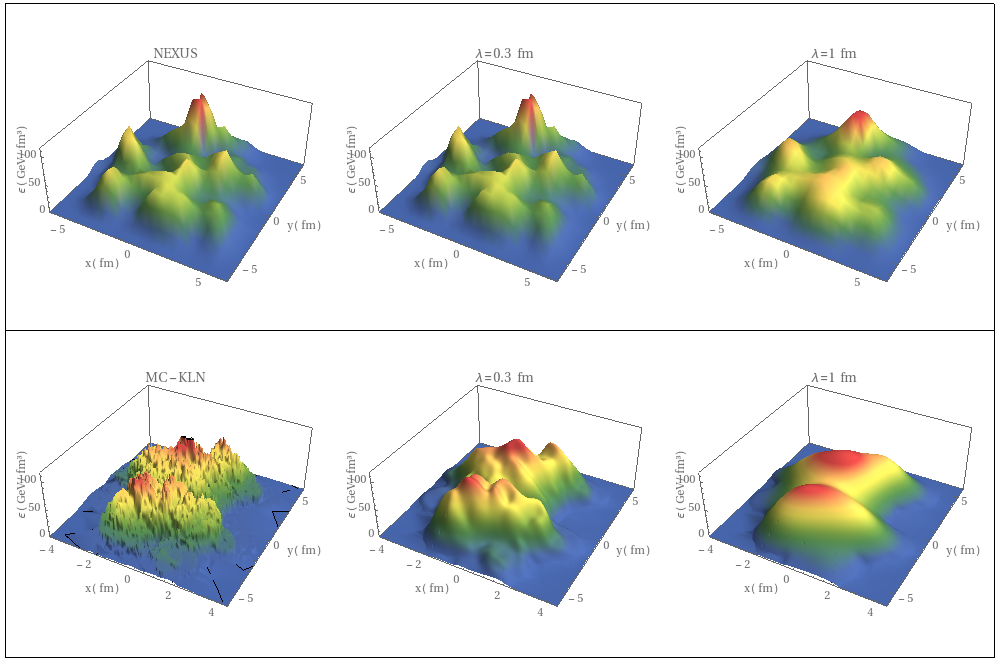}
\caption
{Top: NeXus initial energy density in a midrapidity transverse plane without modification and modified by a cubic spline filter with $\lambda$=0.3 and 1 fm. This corresponds to a central Pb-Pb collision at $\sqrt{s_{NN}}=2.76$ TeV.
{Bottom: MC-KLN initial energy density in a midrapidity transverse plane without modification and  modified by a cubic spline filter with $\lambda$=0.3 and 1 fm. This corresponds to a non-central Pb-Pb collision at $\sqrt{s_{NN}}=2.76$ TeV.}
  \label{fig:IC} 
}
\end{figure*}

In order to investigate the influence of hot spot sizes on observables, 
we modify the initial conditions for each event using a filter. 
The aim is to smooth the energy density profile, such that global properties (as represented by eccentricities $\varepsilon_n$) are kept relatively unchanged, but small-scale structure (quantified by higher cumulants with $m>n$) is different.  This allows to investigate the dependence on the granularity of the initial state.

The  filter we use is based on cubic splines
 and was described (for the two-dimensional case) in \cite{Noronha-Hostler:2015coa}. For completeness we reproduce part of the discussion here.
The idea is that the  transverse energy density value at some point  is determined as a weighted sum of energy density values at fixed points $\vec{r}_\alpha$ around it in the transverse plane, with nearest points contributing more.

 \begin{equation}
   \epsilon(\tau_0, \vec{r}; \lambda)=\sum_{\alpha=1}^{N} \epsilon(\tau_0,\vec{r_\alpha}) W\left(\frac{|\vec{r}-\vec{r_\alpha}|}{\lambda};\lambda\right)
 \end{equation}
 
where W  is given by:

\begin{equation}
  W\left(\frac{|\vec{r}|}{\lambda};\lambda\right)=\frac{10}{7 \pi \lambda^2} f\left(\frac{|\vec{r}|}{\lambda}\right)
\end{equation}
and
\begin{equation}
  f(\xi)=
\left \{
\begin{array}{cc}
  1-\frac{3}{2} \xi^2+\frac{3}{4} \xi^3 & \mbox{if }0\leq \xi < 1\\
  \frac{1}{4} (2-\xi)^3 & \mbox{if }1\leq \xi \leq 2\\
  0 & \mbox{if } \xi > 2
\end{array}  
  \right.
\end{equation}
We note that W is peaked at $\vec{r}=0$, non-negative, invariant under parity and satisfies $\int W\left(\frac{|\vec{r}|}{\lambda};\lambda\right) d\vec{r}=1$
so the integral of $\epsilon(\tau_0, \vec{r}; \lambda)$ on the transverse plane is not modified by a change in $\lambda$.

The advantage of this filter is that it has a compact support and we have a good control of its effect when changing  the value of the parameter $\lambda$.


Figure  \ref{fig:IC}  shows the effect of the filters on a typical event generated with NeXus.
The cubic spline filter with   $\lambda=1$ fm  maintains the locations of the main pikes and valleys but smooth them so that
 their spatial extent increases. The cubic spline filter with  $\lambda=0.3$  fm has little effect as expected since the relevant scale for NeXus initial conditions is the nucleon size.
 The effect of  the cubic spline filter is also illustrated for MC-KLN. Since the initial inhomogeneities occur on a smaller scale, the effect of the filter is stronger for small values of $\lambda$.

\begin{figure}
\includegraphics[width=\linewidth]{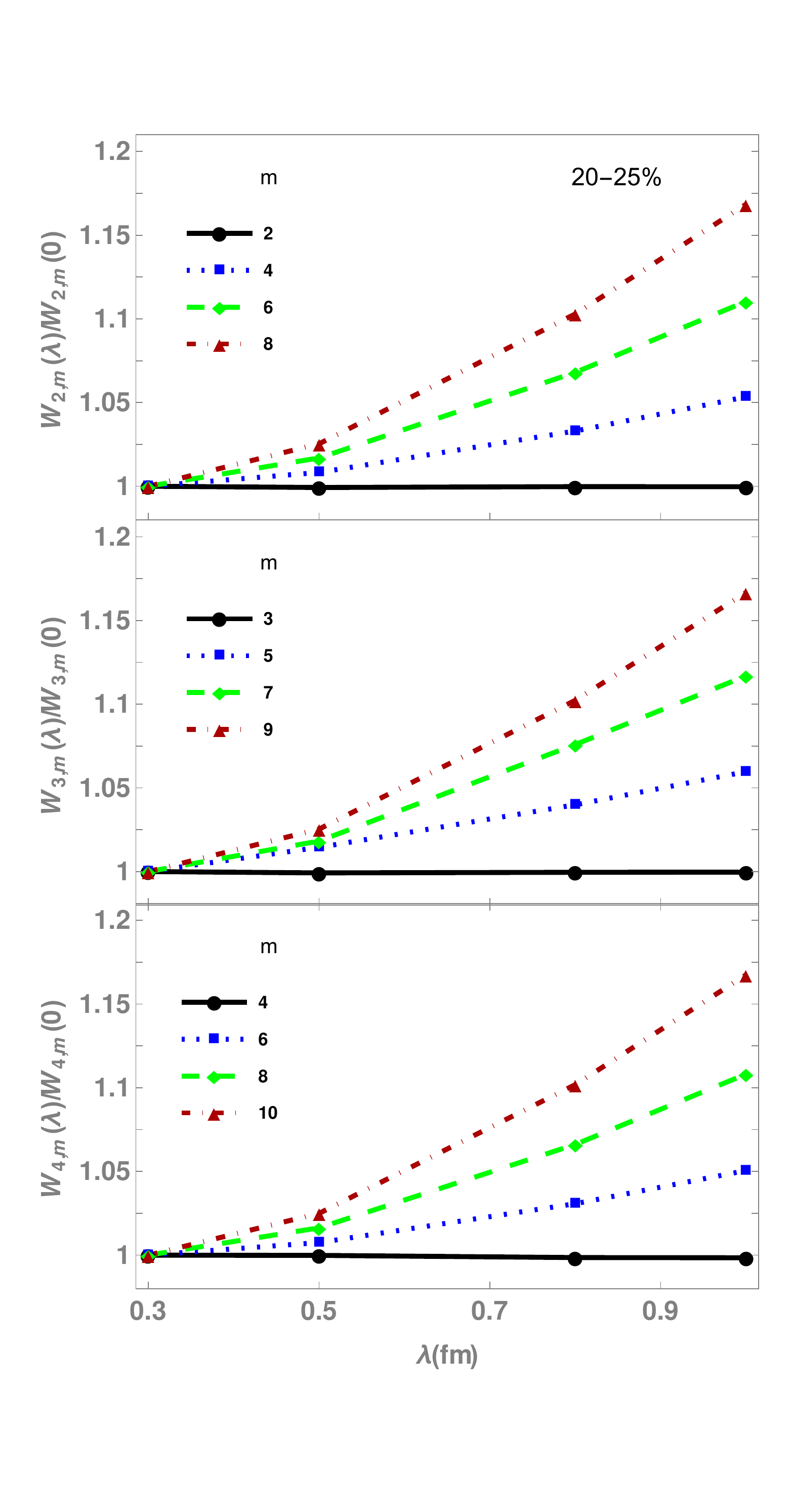}
\caption{\label{fig:Wnm}Cumulants $W_{n,m}$ as a function of smoothing parameter $\lambda$ for NeXus events in the 20-25\% centrality bin. Each cumulant is normalized by its value without smoothing
  $W_{n,m}(0)$.}
\end{figure}

In Fig.~\ref{fig:Wnm}, we show the effect of the smoothing on cumulants $W_{n,m}$ for a set of NeXus events in the 20-25\% centrality bin.  We can see the lowest anisotropic cumulants $W_{n,n}$ are essentially unaffected by smoothing, while higher cumulants depend on the value of the smoothing parameter $\lambda$, with increasing sensitivity for cumulants of larger $m$, as expected.

\begin{figure}
\includegraphics[width=\linewidth]{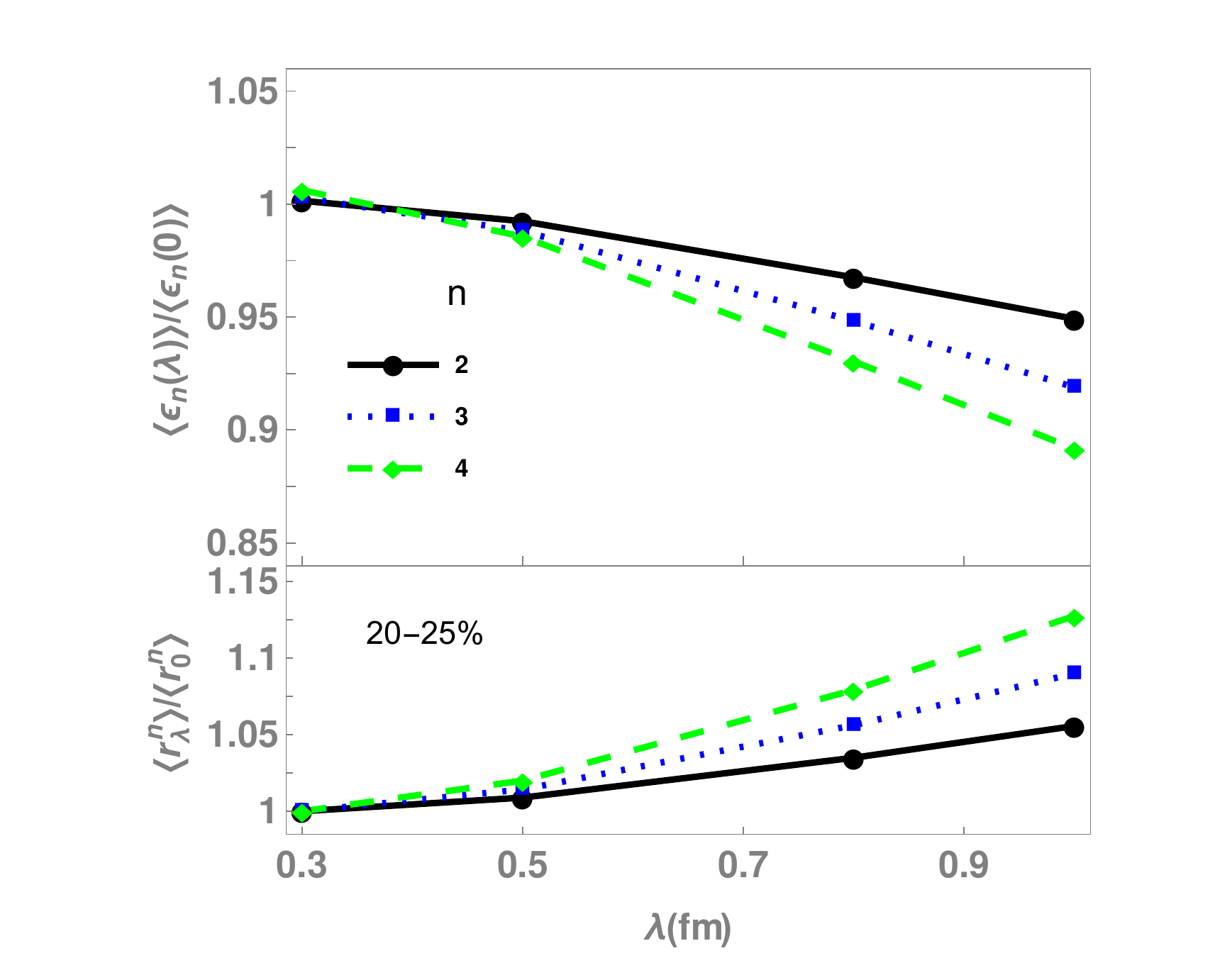}
\caption{\label{fig:ecc} Top: $\langle \epsilon_n\rangle $ as a function of smoothing parameter $\lambda$ for NeXus events in the 20-25\% centrality bin. Each $\langle \epsilon_n\rangle $ is normalized by its value without smoothing $\langle \epsilon_n(0)\rangle $. Bottom: Similar plots for  $\langle r^n\rangle $.}
\end{figure}

Note, however, that the smoothing process does have a small effect on $n=0$ cumulants --- i.e., the size of the system --- as shown in the bottom plot of Fig.~\ref{fig:ecc}.  The average radius of the system increases by $\sim 2.5\%$ when the smoothing parameter is changed from 0 to 1 fm, or $\langle  r^n\rangle\to 1.025^n\langle  r^n\rangle$.  The corresponding eccentricities therefore \textit{decrease} by roughly $n\times$2.5\%.  This is illustrated in the top plot of Fig.~\ref{fig:ecc}.

Any effect that can be explained by this decrease is therefore \textit{not} a dependence on initial state granularity, but only on the well-known dependence on large-scale structure.
For example, if a quantity scales with eccentricity, 
only changes by more than $n$ times the relevant  factor are indicative of a dependence on small scale.
If a ratio of two quantities scaling with eccentricity is considered, 
any change (greater than statistical uncertainty) can be indicative of a dependence on small scale.

Because of this, it is important to use a smoothing procedure that does not significantly increase the size of the system, and this is why in this work we use a filter with compact support.

For MC-KLN, a similar  decrease of eccentricities with $\lambda$ is observed \cite{Noronha-Hostler:2015coa}.

\section{Results for observables}

In this paper, we perform simulations with two codes. Both 
use the Smoothed Particle Hydrodynamics 
Lagrangian algorithm  developed in \cite{testNeX}.

NeXSPheRIO (the first event-by-event code developed for relativistic heavy ion collisions)
 solves the   perfect fluid hydrodynamic equations in 3+1 dimensions.
The initial conditions are obtained event-by-event with the NeXus  generator
\cite{NeXus}.
The equation of state matches lattice data at zero baryonic potential and has a critical point added in a phenomelogical way \cite{eos}.
Isothermal Cooper-Frye freeze out is used with temperatures  chosen in each centrality window to match  observables. At top RHIC energies, this code has successfully reproduced a number of data
\cite{NeXspectra,Andrade06,Andrade08a,Andrade08b,Gardim11,Gardim12a,Takahashi09,Qian12}.
An extension to LHC energies ($\sqrt{s_{NN}} = 2.76$ TeV Pb+Pb collisions)   was developed in \cite{diss_meera} and is used here.
The code was tested against known solutions in \cite{testNeX}.
There is  a 2+1 version of NeXSPheRIO with longitudinal boost invariance that is used here to facilate comparison with the second code described below.

This second code,  v-USPhydro \cite{vuspb,vuspbs}, solves viscous fluid hydrodynamic equations in 2+1 dimensions assuming longitudinal boost invariance. Here it is used to calculate the flow harmonics from  MC-KLN initial conditions  (for $\sqrt{s_{NN} }= 2.76$ TeV Pb+Pb collisions). 
Both (temperature dependent) bulk and shear viscosities can be considered \cite{vuspb,vuspbs}. 
For simplicity's sake, only constant shear viscosity is assumed and adjusted to obtain a reasonable description of LHC data.
 The lattice-based equation of state equation of state S95n-v1 from \cite{eospasi} and an isothermal  Cooper-Frye  freezeout are used (although this choice may affect $\eta/s$ at high energies \cite{Alba:2017hhe}).  v-USPhydro was shown  to reproduce TECHQM test \cite{testtech} as well as both the analytical and semi-analytical radially expanding solutions of Israel-Stewart
hydrodynamics \cite{testvu}.

Note that in the following we also show results from smoothing out IP-Glasma and MC-Glauber initial conditions but do not run them through hydrodynamics.

\subsection{Integrated observables}

As we have seen in Fig. \ref{fig:ecc}, the eccentricities are little affected by the smoothing length for NeXus initial conditions,
changing
at most by $n\times$2.5\%.
Due to the strong event-by-event correlation between final flow and initial eccentricity \eqref{linear}, we  expect a similar change in integrated flow observables.

\begin{figure}[!ht]
\begin{center}
\includegraphics[width=\linewidth]{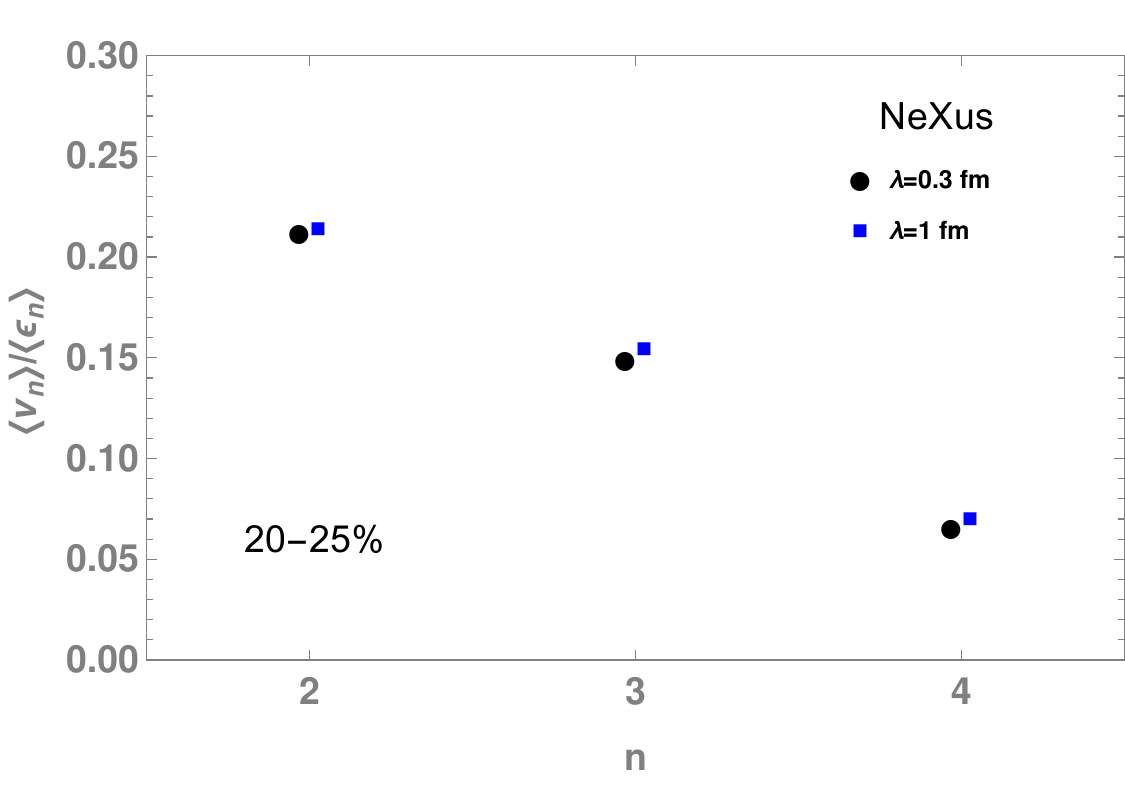}
+\caption{Comparison of original and filtered eccentricity scaled flow harmonics $\langle v_n\rangle/\langle \varepsilon_n\rangle$ for the 20-25\% centrality window.}
\label{fig:vmedio}
\end{center}
\end{figure}

This is indeed the case. 
In  Fig. \ref{fig:vmedio} we show the ratio $\langle v_n\rangle/\langle\varepsilon_n\rangle$ using different smoothing lengths.  Most of the change in integrated $v_n$ is compensated by the change in $\varepsilon_n$, with only a slight residual dependence, in particular for $v_4$, which is known to not follow eccentricity scaling.
There is no indication of a significant dependence on small-scale structure, and instead the results are   determined by the global structure of the initial conditions.

We can make even more precise tests by considering scaled observables that are approximately independent of the small change in system size from our smoothing procedure.  

Therefore, we next consider event-by-event distributions of anisotropic flow $P(v_n)$ \cite{ATLAS12,ATLAS13, ALICE13}.  Equation \eqref{linear} suggests that a uniform change in eccentricity should result in a uniform change in the distribution of $v_n$. If we divide the distribution by the mean, the result $P(v_n/\langle v_n\rangle)$ should then be independent of such a rescaling of eccentricity.   

This is the reason, for example, that scaled distributions of flow coefficients depend little on viscosity, and instead directly probe the initial conditions \cite{Niemi12,Niemi15}.  Because of this, one can immediately see that some models are incompatible with measured data \cite{ATLAS12,ATLAS13} , while others \cite{Schenke13, Niemi15} agree with data (the latter includes the NeXus model used in this work \cite{RHICdist}).   

%

\begin{figure*}[!ht]
\begin{center}
\includegraphics[width=\linewidth]{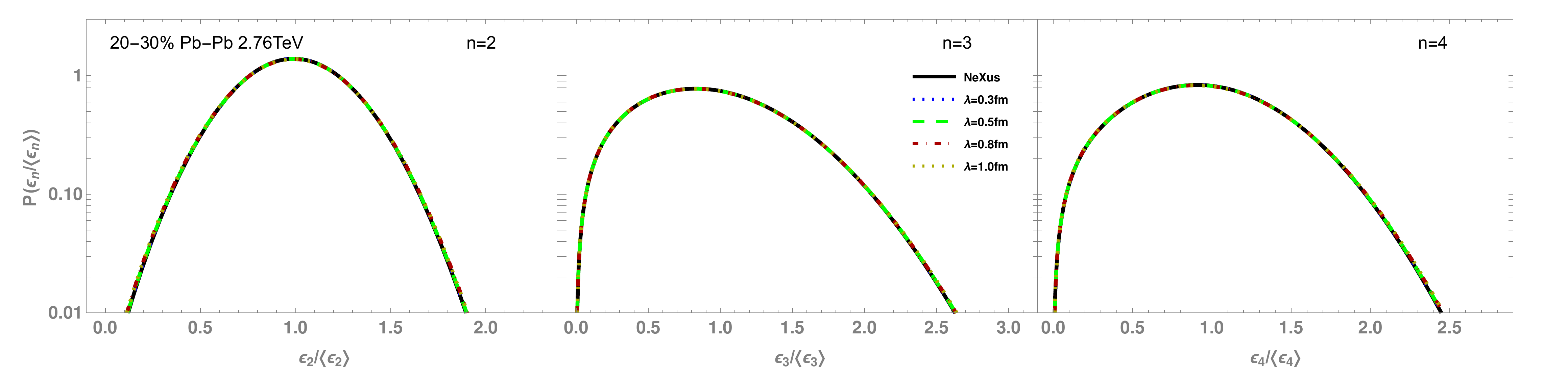}
\caption{Comparison of original and filtered scaled $\epsilon_n$ probability distributions for NeXus initial condition in the 20-30\% windows.}
\label{fig:distem}
\end{center}
\end{figure*}

To study the effect of smoothing,
we first consider the $ P(\epsilon_n/ \langle\epsilon_n\rangle )$ distributions. Results for NeXus initial conditions and ideal hydrodynamics are shown for the 20-25\% centrality window in Fig.~\ref{fig:distem}. No dependence on the value of $\lambda$ is seen, indicating that smoothing essentially corresponds to a uniform rescaling of $\varepsilon_n$ (in turn caused by a uniform scaling of the system size $\langle r^n\rangle$). 

\begin{figure*}[!ht]
\begin{center}
+\includegraphics[width=\linewidth]{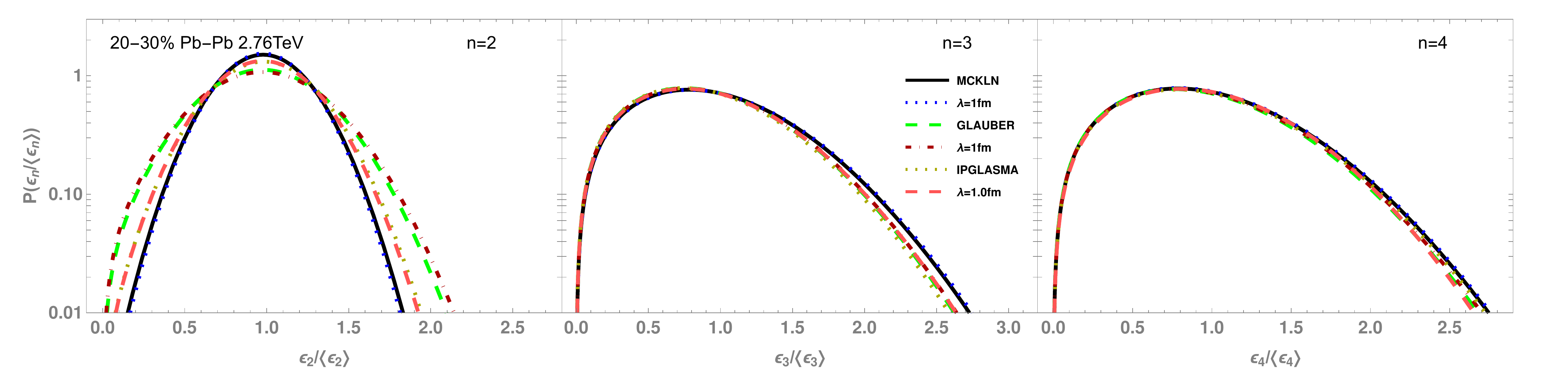}\\
+\caption{Comparison of original and filtered scaled $\epsilon_n$  probability
 distributions for various models of initial conditions in the 20-30\% windows.}
\label{fig:filterecc2dist}
\end{center}
\end{figure*}

One might think that these small differences are due to the fact that in NeXus the typical inhomogeneities have nucleonic size, which is comparable to the maximum $\lambda$ considered. 
More striking results are shown in Fig. \ref{fig:filterecc2dist}. The original and filtered ($\lambda$=1 fm) eccentricity distributions for IP-Glasma  \cite{Schenke12}, MC-KLN \cite{MCKLN} and MC-Glauber \cite{glauber1,glauber2,glauber3}
models were obtained for the 20-30\% centrality window.  One sees that for these  models, the scaled eccentricity distributions are insensitive to the smoothing length below 1 fm.
Similar results hold for other centralities. 

However deviation from Eq.~\eqref{linear} are known to happen; e.g., elliptic flow $v_2$ does not grow perfectly linearly 
with $\epsilon_2$ for non-central collisions \cite{Niemi12,Niemi15,Fu15,Jaki15}.

\begin{figure*}[!ht]
\begin{center}
\includegraphics[width=\linewidth]{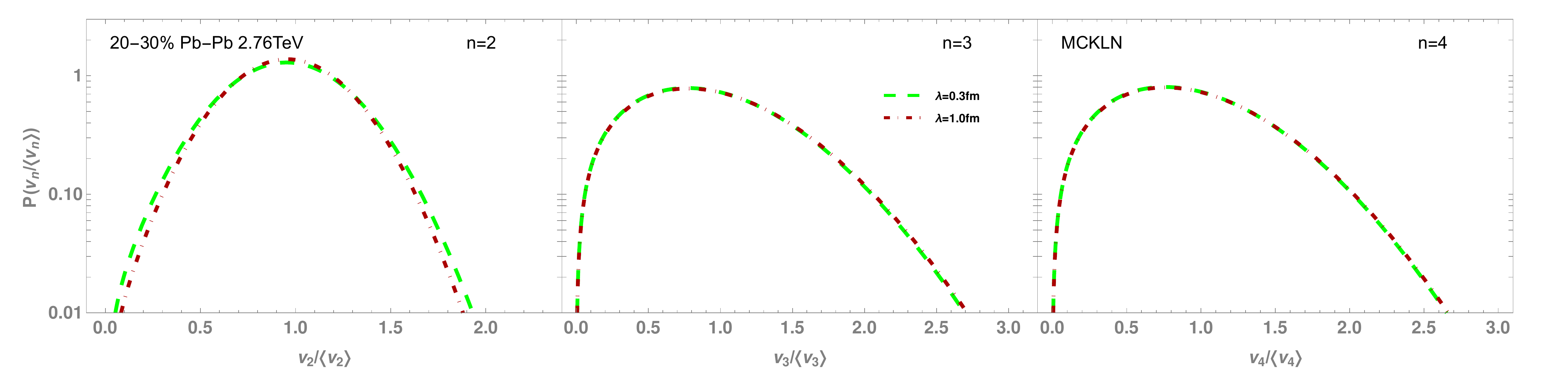}
\caption{Comparison of original and filtered scaled $v_2$ probability distributions for MC-KLN initial conditions}.
\label{fig:filtervndist}
\end{center}
\end{figure*}

Therefore it is important to compute $v_n$ distributions, to determine whether the small deviation from linearity is due to higher cumulants and small-scale structure, or simply a non-linear dependence on eccentricities, so that only global properties are important. The most interesting case is a non-central bin for n=2 since
it was observed 
 \cite{ATLAS13,Renk:2014jja}
that $v_2$ distributions for central collisions as well as $v_3$ and $v_4$ distributions for all centralities do not depend on the details of the initial conditions.
The scaled $v_2$ distributions for the original and filtered initial conditions are compared for MC-KLN in Fig.~\ref{fig:filtervndist} for the 20-30\% centrality window.   
They too are  independent of the value of $\lambda$. 

We conclude that integrated flow $v_n\{2\}$ and event-by-event distributions of anisotropic flow coefficients have little dependence on the smoothing length for the four models considered in this paper, and instead depend only on global features of the initial conditions.
To continue our search for variables that depend on the hot spot size, we note that 
  $v_n$ distributions  contain information only about a single Fourier harmonic $n$. It is then interesting to study  mixed harmonic observables, in particular those that 
are experimentally measurable\cite{aliceSC15,aliceSC16}
or may be obtained at RHIC \cite{gardim17}.

We consider normalized symmetric cumulants:

\begin{equation}NSC(n,m)\stackrel{flow}{=}\frac{\langle v_n^2v_m^2\rangle -\langle v_n^2\rangle \langle v_m^2\rangle }{\langle v_n^2\rangle \langle v_m^2\rangle }
\end{equation}

\begin{figure}[!ht]
\begin{center}
\includegraphics[width=0.5\textwidth]{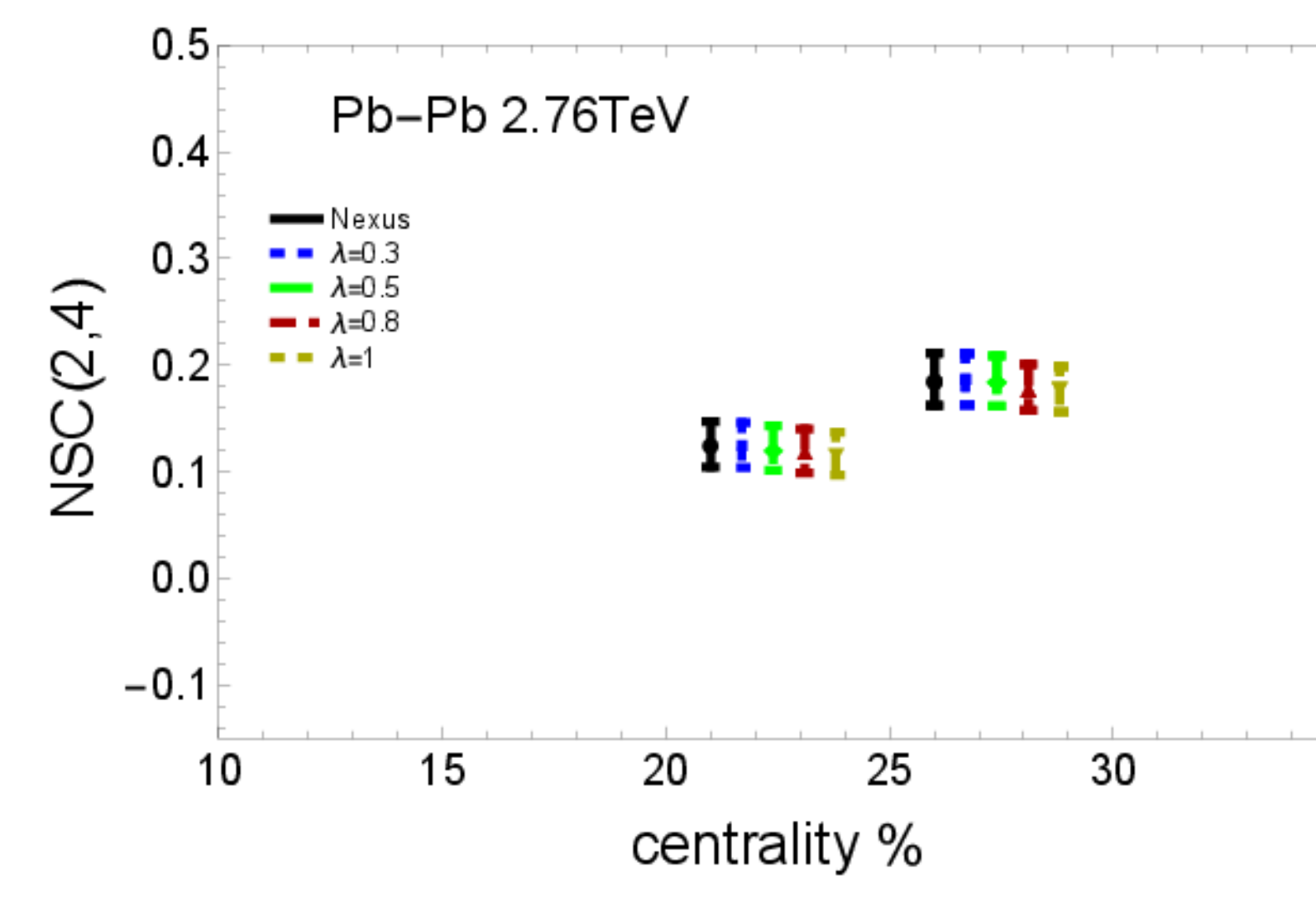} \\
\includegraphics[width=0.5\textwidth]{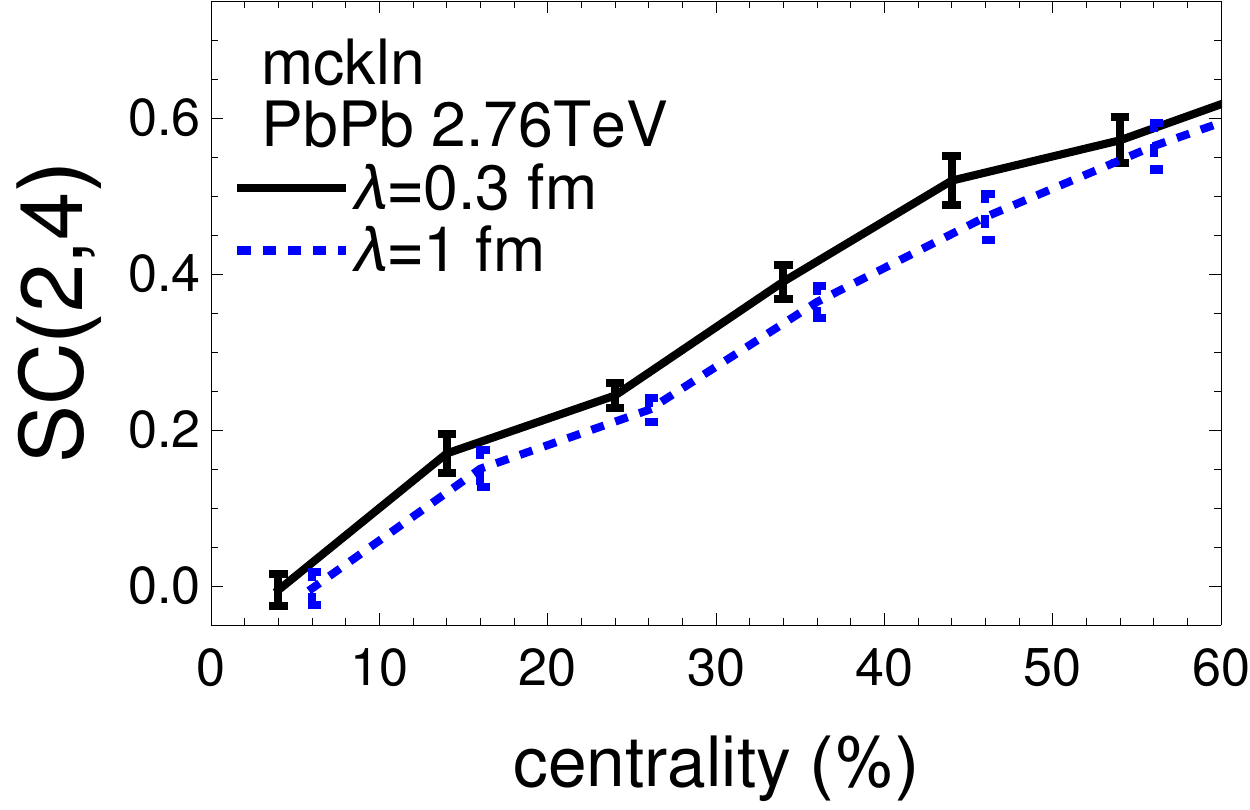} 
\caption{Comparison of  NSC(2,4) for NeXus (top) and MC-KLN (bottom) original and smoothed initial conditions}.
\label{fig:NSC24}
\end{center}
\end{figure}

We note that the connection between these quantities and their equivalent ones computed with eccentricities is not one-to-one.
In Ref.~\cite{jyPCA}, it was argued that NSC(2,3) and NSC(3,4) depend little on the initial conditions while NSC(2,4) does. In figure \ref{fig:NSC24}, one can see that the precise hot spot size does not matter even for NSC(2,4) for NeXus and MC-KLN initial conditions. 

We can go a step farther and consider event plane correlations which mix both magnitude and event planes and have been measured by ATLAS \cite{atlas14}.
However, we also found no dependence on the smoothing length for these observables. 

\subsection{Differential observables}
From $\lambda=0.3-1$ fm no clear evidence of a sensitivity to granularity could be found in $p_T$ integrated observables of all charged particles.
Additional information can be obtained from differential quantities, which we now consider.
Transverse momentum spectra for different hot spot sizes were computed in \cite{Petersen:2010zt,Noronha-Hostler:2015coa} (respectively with URQMD and MC-KLN initial conditions) and exhibit little difference (for hot spot size below 1 fm).
Harmonic flow $v_n(p_T)$'s  were studied in \cite{Petersen:2010zt,Noronha-Hostler:2015coa,RihanHaque:2012wp}, small  changes were found when the hot spot size was varied below 1 fm and  other parameters were held fixed.
In order to find observables that depend on the smoothing length,
we  turn to  another quantity, azimuthal correlations.  The simplest is a pair correlation:

\begin{equation}
  \frac{dN_{pairs}}{d^3p_1 d^3p_2}\propto
  \left[1+\sum\limits_{n=1}^{\infty}2 V_{n\Delta}(p_1,p_2) \cos[n(\phi_1-\phi_2)]
    \right],
\end{equation}

In principle, the  Fourier coefficients $V_{n\Delta}(p_1,p_2)$ depends on two momenta, $p_1$ and $p_2$, which can be varied independently, and the the full matrix 
 has been  measured (e.g. \cite{alice12}). 

%

Since we have already studied the affect of the overall magnitude of anisotropic flow, through momentum integrated measurements, it is convenient to consider a ratio that removes the trivial dependence on $\varepsilon_n$.
To that end, we consider the flow factorization ratio \cite{gardim13}, which was studied in several works \cite{shen13,kozlov14,heinz15}:

\begin{equation}
  r_n(p_1,p_2)=\frac{ V_{n\Delta}(p_1,p_2)}{\sqrt{ V_{n\Delta}(p_1,p_1) V_{n\Delta}(p_2,p_2)}}
  \label{eq:ratio}
 \end{equation} 

This quantity is a good candidate to discriminate smoothing lengths since
 it was shown in \cite{kozlov14} that $r_n$ could be sensitive to the hot spot size but less so to shear viscosity (on this last point see also \cite{heinz15} and \cite{McDonald:2016vlt} for details on bulk viscosity and hadronic rescattering).

\begin{figure}[!ht]
\begin{center}
  \includegraphics[width=0.5\textwidth]{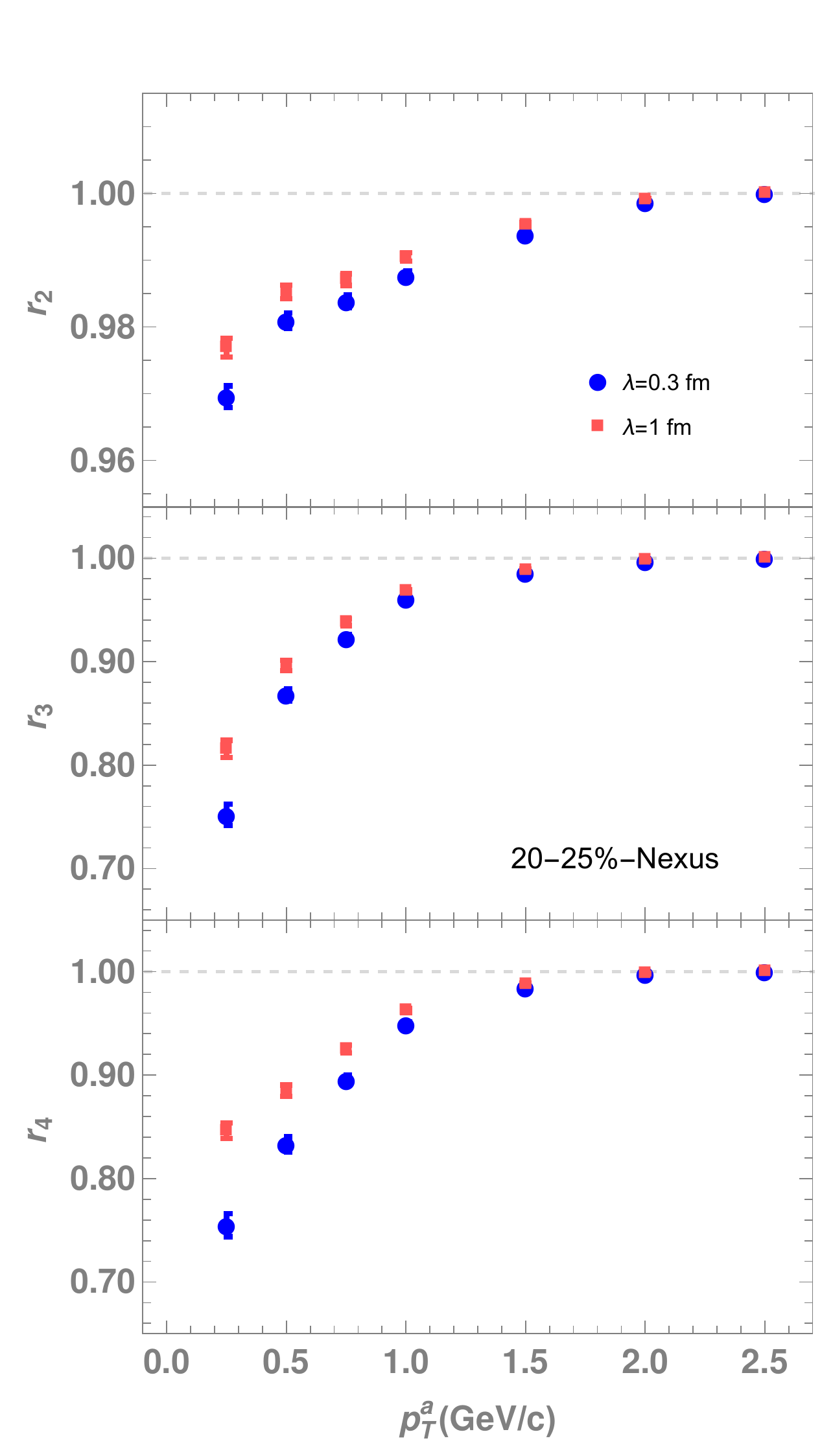} 
  \caption{Flow factorization ratio for NeXus original and smoothed initial conditions in the 20-25\% centrality window.
 }
\label{fig:nexusrn}
\end{center}
\end{figure}

\begin{figure}[!ht]
\begin{center}
  \includegraphics[width=0.5\textwidth]{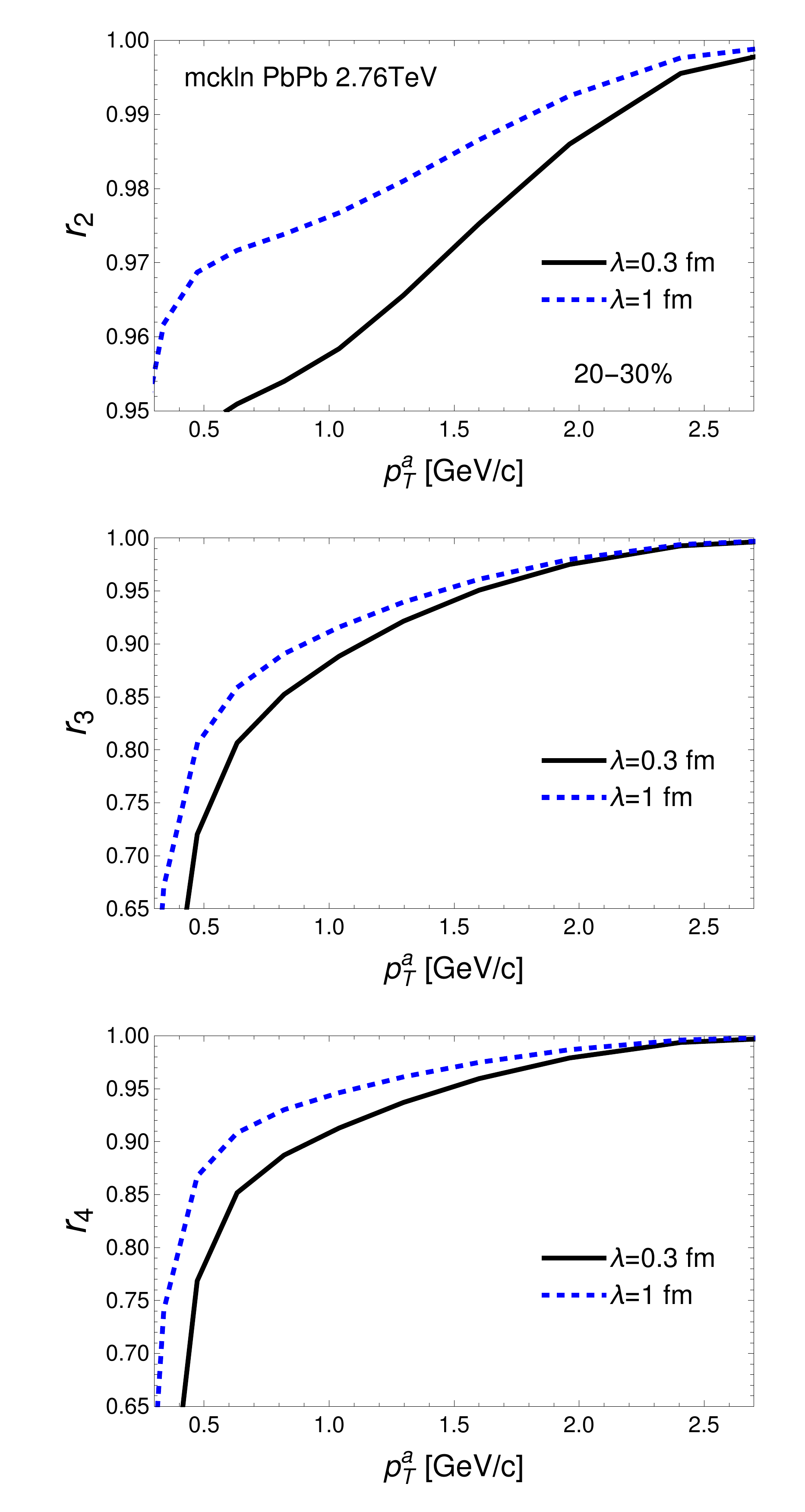} 
\caption{Flow factorization ratio for MC-KLN original and smoothed initial conditions  in the 20-30\% centrality window.
 }
\label{fig:klnrnbr}
\end{center}
\end{figure}

Results for the flow factorization ratios are shown for NeXus and MC-KLN initial conditions in \ref{fig:nexusrn} and
\ref{fig:klnrnbr}. 
Recall that a value $r_n=1$ is obtained in the absence of $p_T$-dependent fluctuations.  The deviation from unity is therefore a measure of the size of such fluctuations.
Thus, we indeed observe a significant dependence on the value of the smoothing scale $\lambda$ on the size of $p_T$-dependent fluctuations, and therefore $r_n$.

In Eq.~(\ref{eq:ratio}), the trivial decrease of the eccentricity with $\lambda$ should approximately cancel between numerator and denumerator. Therefore
the difference (of order 15\% in the most favorable case) is a genuine dependence on smaller scale structures in the initial energy density.

As a final step to search for observables sensitive to hot spot size, we perform a
Principal Component Analysis (PCA). PCA is a method used in statistics to study data that are possibly correlated. It was suggested to apply it to the matrix formed by the coefficients $ V_{n\Delta}(p_1,p_2)$ (with a  different normalization than above) in \cite{jyPCA}. 
A generalization to correlations involving different flow harmonics was proposed in Ref.~\cite{Bozek:2017thv}.
 Further investigations on the connection with initial geometry were done in \cite{teaneyPCA15,teaneyPCA16} and  data from CMS have become recently available \cite{cmsPCA}.

\begin{figure}
\begin{center}
\includegraphics[width=0.5\textwidth]{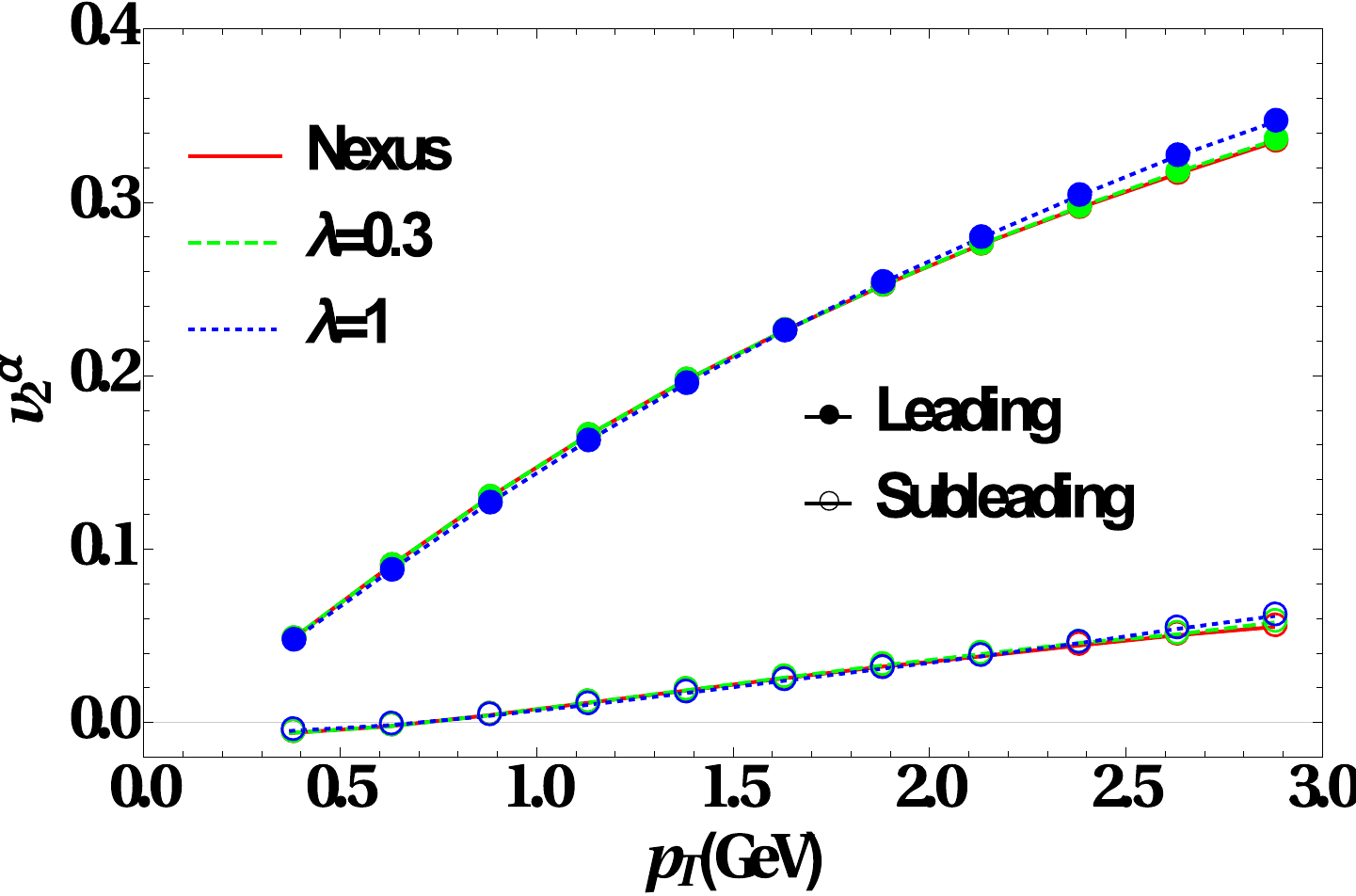} \\
\includegraphics[width=0.5\textwidth]{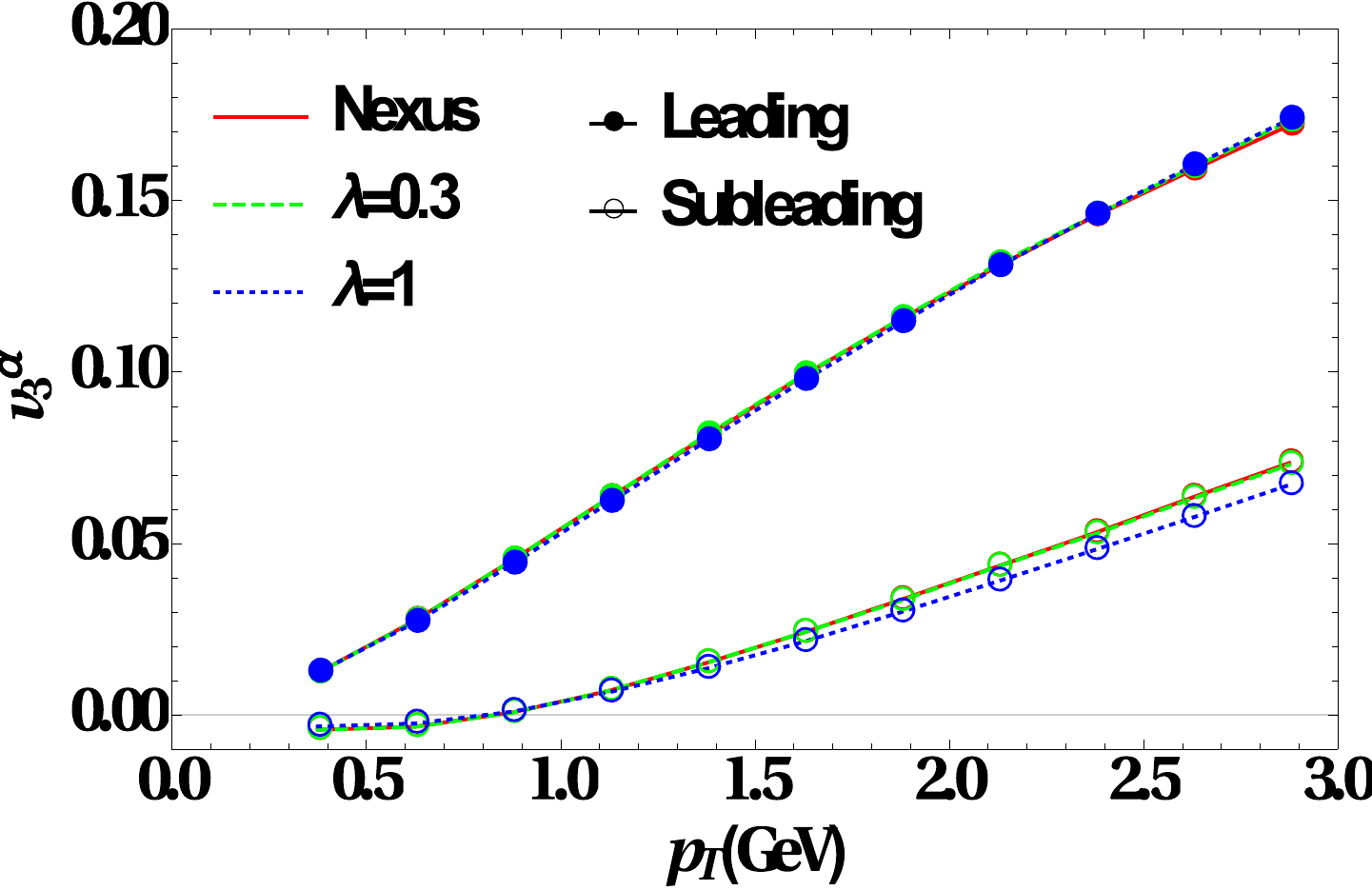} \\
\includegraphics[width=0.5\textwidth]{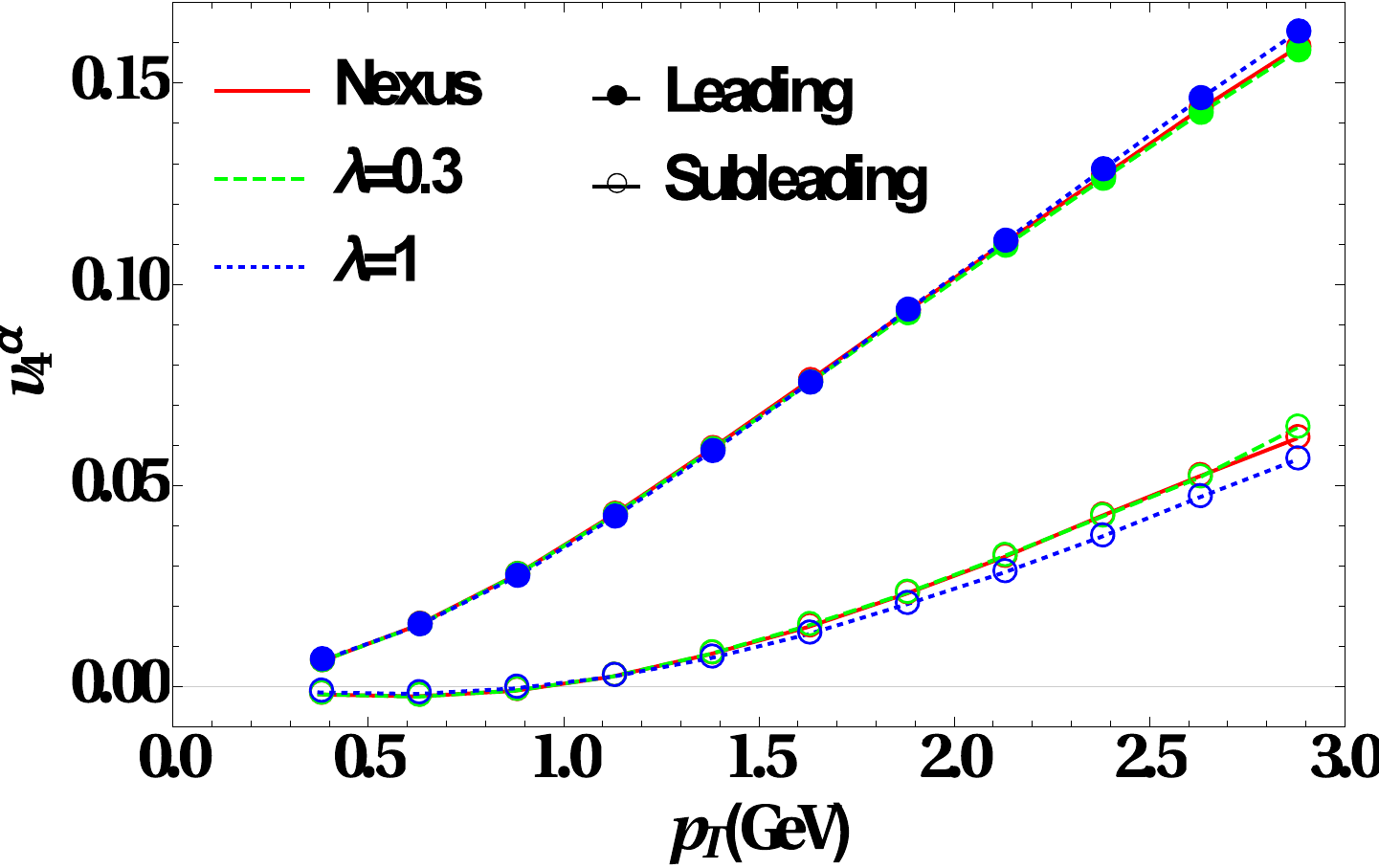} 
  \caption{ Leading and subleading components for NeXus  original and smoothed initial conditions
    in the 25-30\%  centrality class for n=2-4.}
\label{fig:pca}
\end{center}
\end{figure}

We show for n=2-4 the leading principal flow vector (divided by the multiplicity average in the $p_T$ bin) $v_n^{(1)}(p_T)$ in Fig.~\ref{fig:pca} for the 25-30\% centrality bin.
The leading components exhibit a small dependence on the smoothing length, consistent with the change in eccentricity. This is expected since
they contain similar information to $v_n\{2\}(pT)$ \cite{jyPCA}, which are are not very sensitive to hot spot sizes \cite{Petersen:2010zt,Noronha-Hostler:2015coa,RihanHaque:2012wp}.

We also show for n=2-4 the subleading principal flow vector (again divided by the multiplicity average in the $p_T$ bin) $v_n^{(2)}(p_T)$ in fig. \ref{fig:pca}.
They exhibit a small dependence on the smoothing length \footnote{We leave a detailed   comparison of CMS data with 3+1 hydro to a future work}. 
A dependence is not unexpected since  the subleading component is caused by $p_T$-dependent fluctuations (and has a direct relation to factorization breaking) \cite{jyPCA}.  While the effect does not appear to be large, it is of measureable size.

\section{Conclusion}
In this paper,
in order to investigate the influence of hot spot sizes on observables, 
we propose a filter to
modify the initial conditions: it smooths  the energy density profile in such a way such that global properties (as represented by eccentricities $\varepsilon_n$) are kept relatively unchanged, but small-scale structure varies.  
We consider four models of initial conditions (NeXus, MC-Glauber, MC-KLN and IP-Glasma) that have very different size of fluctuations.
We found that when the smoothing length increases from 0.3 to 1 fm, the eccentricities decrease by n times a few percent, due only to the small increase in system size of the smoothing procedure.
Therefore to find a signal of the hot spot sizes in observables scaling with eccentricity, larger changes than that should be seen. In ratio of quantities scaling with eccentricity
any dependence may be genuine.

We note that the focus of this paper has been on small scale structure in large PbPb collisions.  Recently, it has been shown that small systems such as pPb and pp may provide more clues about small scale structure \cite{Noronha-Hostler:2015coa,Welsh:2016siu,Albacete:2017ajt,Giacalone:2017uqx,Moreland:2017kdx}.  We leave a deeper study on small systems for a later work.

We use ideal and viscous hydrodynamics and  compute a range of observables. 
We find that integrated $v_n$ values, scaled $v_n$ distributions, normalized symmetric cumulants, event-plane correlations, leading component in a Principal Component Analysis (and therefore $v_n(p_T)$)
do not have a significant dependence on small-scale structure. However the factorization breaking ratio and subleading principle components exhibit non-trivial dependence on the smoothing length.
Since the factorization breaking ratio depends little on viscosity, it is the best observable we found to discriminate models that have different fluctuation sizes.

\section{Acknowledgements}
We  thank  J.-Y.~Ollitrault for very helpful discussions on the PCA method. 
J.N.H acknowledges the use of the Maxwell Cluster and the advanced support from the Center of Advanced Computing and Data Systems at the University of Houston.
F.G.~acknowledges  support  from
Funda\c{c}\~ao de Amparo \`a Pesquisa do Estado de S\~ao Paulo
(FAPESP  grants  2015/00011-8, 2015/50438-8,  2016/03274-2), USP-COFECUB (grant Uc Ph 160-16 (2015/13) ) and
Conselho Nacional de Desenvolvimento Cient\'{\i}fico e Tecnol\'ogico (CNPq grant 310141/2016-8).  F.G.G. was supported by Conselho Nacional de Desenvolvimento Cient\'{\i}fico  e  Tecnol\'ogico  (CNPq grant 449694/2014-3 and 312203/2015-2)  and
FAPEMIG (grant APQ-02107-16.
F.G.G.~and P.M. acknowledge computing time provided on the PdCluster made available by L.F.R. Turci and E. Aguilar (Universidade de Federal de Alfenas/Poços de Caldas).
 P.I.~thanks support from Coordena\c{c}\~ao de Aperfei\c{c}oamento de Pessoal de N\'{\i}vel Superior (CAPES).
M.L.~acknowledges support from FAPESP projects 2016/24029-6  and 2017/05685-2, and project INCT-FNA Proc.~No.~464898/2014-5

\bibliography{bibli}

\end{document}